\documentclass[12pt]{article}
\usepackage{amssymb}
\usepackage{mathrsfs}
\usepackage{amsmath,amssymb,amsthm,amscd}
\usepackage{epsf,amsfonts,hyperref}
\usepackage{color}

\usepackage{cite}
\usepackage{graphicx}

\input epsf.sty
\makeatletter
\def\@biblabel#1{#1.}
\makeatother

\newcommand{\be}{\begin{equation}}
\newcommand{\ee}{\end{equation}}
\newcommand{\ba}{\begin{aligned}}
\newcommand{\ea}{\end{aligned}}

\renewcommand{\appendix}[1]{
    \addtocounter{section}{1}
    \setcounter{equation}{0}
    \renewcommand{\thesection}{\Alph{section}}
    \section*{Appendix \thesection\protect\indent #1}
    \addcontentsline{toc}{section}{Appendix \thesection\ \ \ #1}
}
\newcommand\encadremath[1]{\vbox{\hrule\hbox{\vrule\kern8pt
\vbox{\kern8pt \hbox{$\displaystyle #1$}\kern8pt}
\kern8pt\vrule}\hrule}}
\def\enca#1{\vbox{\hrule\hbox{
\vrule\kern8pt\vbox{\kern8pt \hbox{$\displaystyle #1$} \kern8pt}
\kern8pt\vrule}\hrule}}

\newcommand\figureframex[3]{
\begin{figure}[bth]
\hrule\hbox{\vrule\kern8pt \vbox{\kern8pt \vbox{
\begin{center}
{\mbox{\epsfxsize=#1.truecm\epsfbox{#2}}}
\end{center}
\caption{#3} }\kern8pt} \kern8pt\vrule}\hrule
\end{figure}
}
\newcommand\figureframey[3]{
\begin{figure}[bth]
\hrule\hbox{\vrule\kern8pt \vbox{\kern8pt \vbox{
\begin{center}
{\mbox{\epsfysize=#1.truecm\epsfbox{#2}}}
\end{center}
\caption{#3} }\kern8pt} \kern8pt\vrule}\hrule
\end{figure}
}

\newcommand{\beq}{\begin{equation}}
\newcommand{\eeq}{\end{equation}}
\newcommand{\bea}{\begin{eqnarray}}
\newcommand{\eea}{\end{eqnarray}}

\makeatother

\newif\iffigs\figsfalse
\figstrue

\iffigs
  \input epsf
\else
 \fi

\begin{document}

\begin{titlepage}
\vskip .6cm
\centerline{\Large \bf New Algebraic Structures from Hermitian}
\vspace*{1.5ex}
\centerline{\Large \bf One-Matrix Model}

\medskip

\vspace*{4.0ex}

\centerline{\large \rm XIANG-MAO DING\footnote{xmding@amss.ac.cn,
supported by NSFC No. 11375258}, YUPING
LI\footnote{liyuping@amss.ac.cn, corresponding author} and LINGXIAN
MENG\footnote{menglingxian@amss.ac.cn} }

\vspace*{4.0ex}

\centerline{ \rm~Institute of Applied Mathematics}
\centerline{Academy of Mathematics and Systems Science}
\centerline{Chinese Academy of Science, Beijing 100190; P.R.China}

\vspace*{10ex}

\noindent {\bf Abstract.} Virasoro constraint is the operator algebra version of one-loop equation for a Hermitian one-matrix model, and it plays an important role in solving the model. We construct the realization of the Virasoro constraint from the Conformal Field Theory (CFT) method.  From multi-loop equations of the one-matrix model, we get a more general constraint. It can be expressed in terms of the operator algebras, which is the Virasoro subalgebra with extra parameters. In this sense, we named as generalized Virasoro constraint. We enlarge this algebra with central extension, this is a new kind of algebra, and the usual Virasoro algebra is its subalgebra. And we give a bosonic realization of its subalgebra.
\bigskip



\end{titlepage}
\vfill \eject

\section{Introduction }

Matrix models play important roles in physics and mathematics. They unify many seemingly unrelated disciplines, including such as integrable systems, conformal field theories, topological expansion, etc \cite{M.L.Metha,P.Di.Francesco,A.Morozov,I.Kostov}. They are involved in condensed matter physics \cite{Les.Houches,T.Guhr}, statistical physics \cite{F.David,D.Gross}, high energy physics \cite{P.Di.Francesco,T.Banks}, and even they are linked to the probability theory \cite{M.L.Metha}, the moduli space of curves \cite{M.Kontsevich,R.Penner} and the Riemann conjecture \cite{B.Eynard4}. For more interesting topics related to matrix models, please see \cite{E.Brezin} for a nice review.

The partition function of matrix models can be considered as generating functions of discrete surfaces \cite{B.Eynard7}. It is well known \cite{P.Di.Francesco} that the free energy and the correlation functions of matrix models can be expanded with respect to the parameter  $\frac{1}{N^{2}}$, and it is called as topological expansions. In general, these partition functions are not a convergent integrals, and they are understood as formal series in the coefficients of the potential. The relevant matrix models are often known as formal matrix model.

There are different methods for solving matrix models (both convergent and formal). Among them, the topological recursion is a new and effective way to solve the formal matrix integral. The topological recursion is based on the topological expansion of the formal matrix integral. B.Eynard and his collaborators give the recursive formulas in \cite{L.Chekhov,L.Chekhov1,B.Eynard1,B.Eynard5,B.Eynard6}. If the size of the matrix is finite, there is a genus expansion\cite{A.Alexandrov}. The footstone of Eynard's formula is  loop-equations. The large $N$ limit of the one-loop equation provides the spectral curve, while the multi-loop equations give the recursive formula. The free energy and the correlation functions can be expressed as function and multi-differentials on the spectral curve, respectively. Thereafter, B.Eynard and N.Orantin extend the results far away from the matrix models, and they construct the free energy and the correlation functions on arbitrary curves. The key point is that these functions are regarded as symplectic invariants of the curves \cite{B.Eynard2,B.Eynard3}.

In the one-matrix model case, the one-loop equation can be expressed as operators acting on the free energy. As the operators satisfy a Virasoro algebra relation, that they are called as the Virasoro constraint\cite{A.Mironov,F.David1,J.Ambjern1,H.Itoyama,A.Gerasimov}. They can be seen as annihilation operators of a given realization of the Virasoro algebra. In this paper, we will construct a realization of the Virasoro algebra with the CFT method, for which the annihilation part is exactly the Virasoro constraint. Like the one-loop equation, the multi-loop equations can also be expressed as operators acting on the free energy of one-matrix model. We treat these operators as a generalized Virasoro constraint. To investigate the relations of the generalized Virasoro constraint is the goal of this paper. We construct a new operator algebra with the current operators and the Virasoro operators, such that the generalized Virasoro constraint belongs to this new operator algebra.

The paper is organized as follows. In section 2, we review the definition of formal matrix integral and the loop equations of one-hermitian matrix model. In section 3, we derive the Virasoro constraint from the one-loop equation. We give a realization of Virasoro algebra which annihilation part is exactly the Virasoro constraint. In section 4, we get the generalized Virasoro constraint from multi-loop equations, and give the algebraic relation of them. A new Lie algebra is given from these relations, and a bosonic kind realization for the generalized Virasoro constraint is obtained in this section.

\section{Matrix model and the Loop Equations}
\subsection{Formal Hermitian Matrix Model}
Consider a formal Hermitian matrix integral, the partition function
\begin{equation}\label{lyp1}
Z=\int_{H_{N\times N}}dMe^{-N\text{tr}(V(M))},
\end{equation}
where $M$ is a $N\times N$ Hermitian matrix, $dM$ is the product of Lebesgue measures of all real components of $M$. The potential $V(M)$ is a polynomial of degree $d+1\geq2$. For a function $f(M)$, the average $\langle f(M)\rangle$ is
\begin{equation}\label{lyp54}
\langle f(M)\rangle=\dfrac{1}{Z}\int dMf(M)e^{-N\text{tr}(V(M))}.
\end{equation}
Let $f_{1}(M),\cdots,f_{l}(M)$ be functions of Hermitian matrix $M$, the connected part or cumulant of the average $\langle f_{1}(M)\cdots f_{k}(M)\rangle_{c}$ is defined as
\begin{equation}\label{lyp55}
\begin{array}{llllllll}
\langle f_{1}(M)\rangle_{c}=\langle f_{1}(M)\rangle,\\
\langle f_{1}(M)f_{2}(M)\rangle_{c}=\langle f_{1}(M)f_{2}(M)\rangle-\langle f_{1}(M)\rangle\langle f_{2}(M)\rangle,\\
\langle f_{1}(M)f_{2}(M)f_{3}(M)\rangle_{c}=\langle f_{1}(M)f_{2}(M)f_{3}(M)\rangle-\langle f_{1}(M)f_{2}(M)\rangle_{c}\langle f_{3}(M)\rangle_{c},\\
-\langle f_{1}(M)f_{3}(M)\rangle_{c}\langle f_{2}(M)\rangle_{c}-\langle f_{2}(M)f_{3}(M)\rangle_{c}\langle f_{1}(M)\rangle_{c}
-\langle f_{1}(M)\rangle_{c}\langle f_{2}(M)\rangle_{c}\langle f_{3}(M)\rangle_{c},\\
\cdots\cdots.
\end{array}
\end{equation}
The connected part or cumulant of the correlation functions are
\begin{equation}\label{lyp2}
\overline{W}_{k}(x_{1},x_{2},\cdots,x_{k}):=N^{k-2}\left\langle \text{tr}\dfrac{1}{x_{1}-M}\text{tr}\dfrac{1}{x_{2}-M}\cdots \text{tr}\dfrac{1}{x_{k}-M}\right\rangle_{c},
\end{equation}
for $k\geq1$. Furthermore, we introduce the average $\overline{U_{k}}(x_{1};x_{2},\cdots,x_{k})$ that will be used in the following parts of the paper
\begin{equation}\label{lyp3}
\overline{U_{k}}(x_{1};x_{2},\cdots,x_{k}):=N^{k-2}\left\langle \text{tr}\dfrac{V(x_1)-V(M)}{x_{1}-M}\text{tr}\dfrac{1}{x_{2}-M}\cdots\text{tr}\dfrac{1}{x_{k}-M}\right\rangle_{c}.
\end{equation}

When $Z$ is considered as a formal generating function, it is well known that the correlation functions $\overline{W}_{k}(x_{1},x_{2},\cdots,x_{k})$ can expanded with respect to the parameter $\frac{1}{N^{2}}$, that is the so called topological expansion \cite{B.Eynard7}
\begin{equation}\label{lyp4}
\overline{W}_{k}(x_{1},x_{2},\cdots,x_{k}):=\sum_{g=0}^{\infty}N^{-2g}W_{k}^{(g)}(x_{1},x_{2},\cdots,x_{k}),
\end{equation}
Similarly, $\overline{U}_{k}(x_{1},x_{2},\cdots,x_{k})$ have the structure:
\begin{equation}\label{lyp13}
\overline{U}_{k}(x_{1};x_{2},\cdots,x_{k}):=\sum_{g=0}^{\infty}N^{-2g}U_{k}^{(g)}(x_{1};x_{2},\cdots,x_{k}).
\end{equation}
The free energy $\mathcal {F}$ of the formal Hermitian matrix integral \eqref{lyp1} is
\begin{equation}\label{lyp5}
\mathcal {F}:=\log Z.
\end{equation}
The free energy $\mathcal {F}$ can also be expressed as:
\begin{equation}\label{lyp6}
\mathcal {F}=\log Z:=\sum_{g=1}^{\infty}N^{2-2g}\mathcal {F}_{g}.
\end{equation}
This is the so called topological expansion of the free energy. The topological recursion is base on this expression. If the potential $V(x)=\sum\limits_{k\geq0}t_{k}x^{k}$, then the action of loop insertion operator \cite{J.Ambjern}
\begin{equation}\label{lyp7}
\dfrac{\partial}{\partial V(x)}:=-\sum_{k=1}^{\infty}\dfrac{1}{x^{k+1}}\dfrac{\partial}{\partial t_{k}},
\end{equation}
on the correlation function $\overline{W}_{k}(x_{1},\cdots,x_{k})$ is \cite{G.Akemann}
\begin{equation}\label{lyp8}
\begin{array}{lll}
\overline{W}_{k+1}(x_{1},x_{2},\cdots,x_{k+1})=\dfrac{\partial}{\partial V(x_{k+1})}\overline{W}_{k}(x_{1},x_{2},\cdots,x_{k})\\
=\dfrac{\partial}{\partial V(x_{1})}\dfrac{\partial}{\partial V(x_{2})}\cdots\dfrac{\partial}{\partial V(x_{k})}\dfrac{\partial}{\partial V(x_{k+1})}\mathcal {F}.
\end{array}
\end{equation}

\subsection{Loop Equations}
Loop equation is the name given to the Schwinger-Dyson equation in the context of random matrices. It is usually got by request of invariance for the matrix integral \eqref{lyp1} under the change of variables $M\rightarrow M+\epsilon\delta M$:
\begin{equation}\label{lyp9}
\delta M=\dfrac{1}{x_{1}-M}.
\end{equation}
From the linear part of $\epsilon$, we get the one-loop equation:
\begin{equation}\label{lyp10}
\overline{W}_{1}^{2}(x_{1})+\dfrac{1}{N^{2}}\overline{W}_{2}(x_{1},x_{1})=V^{\prime}(x_{1})
\overline{W}_{1}-\overline{U}_{1}(x_{1}).
\end{equation}
Applying the loop insertion operator \eqref{lyp7} to equation \eqref{lyp10}, we get the multi-loop equation:
\begin{equation}\label{lyp11}
\begin{array}{llllllll}
&&2\overline{W}_{1}(x_{1})\overline{W}_{k}(x_{1},\cdots,x_{k})+\dfrac{1}{N^{2}}
\overline{W}_{k+1}(x_{1},x_{1},x_{2}\cdots,x_{k})\\
&+&\sum\limits_{j=1}^{k-2}\sum\limits_{I\in K_{j}}\overline{W}_{j}(x_{1},x_{I})\overline{W}_{k-j}(x_{1},x_{K-I})\\
&+&\sum\limits_{j=2}^{k}\dfrac{\partial}{\partial x_{j}}\dfrac{\overline{W}_{k-1}(x_{2},\cdots,x_{j},\cdots,x_{k})-
\overline{W}_{k-1}(x_{2},\cdots,x_{1},\cdots,x_{k})}{x_{j}-x_{1}}\\
&=&V^{\prime}(x_{1})\overline{W}_{k}(x_{1},\cdots,x_{k})-\overline{U}_{k}(x_{1};x_{2},\cdots,x_{k}).
\end{array}
\end{equation}
And $K=\{2,\cdots,k\}$, and $K_{j}=\{I\subset K|\# I=j\}$ for any $j\leq k-1$ in the equation, and $x_{I}:=x_{i_{1}},x_{i_{2}},\cdots,x_{i_{j}}$,if $I=\{i_{1},i_{2},\cdots,i_{j}\}$.
For equation \eqref{lyp10}, the leading term in $\dfrac{1}{N^{2}}$ expansion is
\begin{equation}\label{lyp12}
W_{1}^{(0)}(x_{1})^{2}=V^{\prime}(x_{1})W_{1}^{(0)}(x_{1})-U_{1}^{(0)}(x_{1}).
\end{equation}
Denoting $Y(x_{1})=V^{\prime}(x_{1})-2W_{1}^{(0)}(x_{1})$, from equation \eqref{lyp12}, we get an algebraic equation about $x_{1},Y_(x_{1})$
\begin{equation}\label{lyp15}
Y(x_{1})^2=V^{\prime}(x_{1})-4U_{1}^{(0)}(x_{1}).
\end{equation}
The explicit expression of $Y(x_{1})$ is
\begin{equation}\label{lyp14}
Y(x_{1})=\sqrt{V^{\prime}(x_{1})-4U_{1}^{(0)}(x_{1})}.
\end{equation}
From equation \eqref{lyp15}, one could get another solution $Y(x_{1})=-\sqrt{V^{\prime}(x_{1})-4U_{1}^{(0)}(x_{1})}$. But $W_{1}^{(0)}(x_{1})=\frac{1}{x_{1}}+O(x_{1}^{-2})$ as $x_{1}\rightarrow \infty$, so equation \eqref{lyp14} is indeed the solution we wanted.

Equation \eqref{lyp15} is the spectral curve for one-hermitian matrix model \eqref{lyp1}. B.Eynard and N.Orantin have developed a method to compute the correlation functions and the free energy of matrix models \cite{B.Eynard1,L.Chekhov} based on expressions \eqref{lyp4}, \eqref{lyp13} and \eqref{lyp6}. They expressed the correlation functions as multiple differentials on the spectral curve, and could compute them recursively, that is the so called E-O formula\cite{B.Eynard2,B.Eynard3}. Equation \eqref{lyp15} (i.e. \eqref{lyp10}) and equation \eqref{lyp11} play important roles in their formula.

\section{Virasoro Constraint}
One of the important properties of matrix models is that, the free energy satisfies a Virasoro constraint. For one-hermitian matrix model, the Virasoro constraint is the operator representation of the one-loop equation. In this section, firstly, we reexpress equation \eqref{lyp10} as a Virasoro constraint, and then extend it in a way such that it forms a whole Virasoro algebra. In the next section, we will give a more general Virasoro constraint from equation \eqref{lyp11} .

From the definition of correlation functions $\overline{W}_{k}(x_{1},\cdots,x_{k})$, we can expand them with respect to the parameters $x_{1},x_{2},\cdots,x_{k}$ in the following way
\begin{equation}\label{lyp16}
\begin{array}{llllll}
\overline{W}_{k}\left(x_{1},\cdots,x_{k}\right)&=&N^{k-2}\left\langle tr\dfrac{1}{x_{1}-M}\cdots tr\dfrac{1}{x_{k}-M}\right\rangle_{c}\\
&=&\sum\limits_{l_{1}=0}^{\infty}\sum\limits_{l_{2}=0}^{\infty}\cdots
\sum\limits_{l_{k}=0}^{\infty}x_{1}^{-l_{1}-1}x_{2}^{-l_{2}-1}\cdots x_{k}^{-l_{k}-1}\left\langle trM^{l_{1}}trM^{l_{2}}\cdots trM^{l_{k}}\right\rangle_{c}.
\end{array}
\end{equation}
Using equation \eqref{lyp16} and the expression of potential $V(x)$, we can reform equation \eqref{lyp10} as:
\begin{equation}\label{lyp17}
\begin{array}{lll}
\sum\limits_{n=-1}^{\infty}x_{1}^{-n-2}\dfrac{1}{N^{2}}\left\{\sum_{k=0}^{n}\left\langle trM^{k}trM^{n-k}\right\rangle-N\sum_{k=1}^{\infty}kt_{k}\left\langle trM^{k+n}\right\rangle\right\}=0.
\end{array}
\end{equation}
In fact, from the definition of matrix integral \eqref{lyp1} and the expression of the potential $V(x)$, equation \eqref{lyp17} can be reexpressed as operators acting on the partition function $Z$,
\begin{equation}\label{lyp18}
\sum_{n=-1}^{\infty}x_{1}^{-n-2}\cdot\dfrac{1}{N^{2}\cdot Z}\left\{\dfrac{1}{N^{2}}\sum_{k=0}^{n}\dfrac{\partial^{2}}{\partial t_{k}\partial t_{n-k}}+\sum_{k=1}^{\infty}kt_{k}\dfrac{\partial}{\partial t_{k+n}}\right\}\cdot Z=0.
\end{equation}
If we define the operators $\mathscr{L}_{n}$ as
\begin{equation}\label{lyp19}
\mathscr{L}_{n}=\sum_{k=1}^{\infty}kt_{k}\dfrac{\partial}{\partial t_{k+n}}+\dfrac{1}{N^{2}}\sum_{k=0}^{n}\dfrac{\partial^{2}}{\partial t_{k}\partial t_{n-k}},\ \ n\geq-1.
\end{equation}
equation \eqref{lyp18} is identical with
\begin{equation}\label{lyp56}
\sum\limits_{n=-1}^{\infty}x_{1}^{-n-2}\mathscr{L}_{n}\cdot Z=0,
\end{equation}
and the operators $\mathscr{L}_{m}$ subject to the relation
\begin{equation}\label{lyp20}
\left[\mathscr{L}_{m}, \mathscr{L}_{n}\right]=\left(m-n\right)\mathscr{L}_{m+n};\  m,n\geq-1.
\end{equation}
In equation \eqref{lyp56}, $x_{1}$ is a variable, so the coefficient of the power of $x_{1}^{-1}$ is zero, i.e.
\begin{equation}\label{lyp57}
\mathscr{L}_{n}\cdot Z=0,\ \ \ n\leq-1.
\end{equation}
The set of operators $\{\mathscr{L}_{n}, n\geq-1\}$ is a subalgebra of Virasoro algebra, and equation \eqref{lyp57} is usually called as a Virasoro constraint of the one-hermitian matrix model.

The partition function of one-hermitian matrix model is a singular vector of the Virasoro algebra. We can regard the partition function $Z$ (i.e.\eqref{lyp1}) as a highest weight state of the Virasoro algebra. We now give a realization of the Virasoro algebra with the CFT method \cite{R.Blumenhagen,S.Kharchev,A.Marshakov}, such that the annihilation subalgebra is exact the $\{\mathscr{L}_{n}, n\geq-1\}$. Setting
\begin{equation}\label{lyp21}
j_{n}=\left\{\begin{array}{cc} -Nnt_{-n}, \ \ n<0;\\
\dfrac{k}{N}\dfrac{\partial}{\partial t_{n}}, \ \ n\geq 0;
\end{array}\right.,
\end{equation}
then the operators $j_{n}$ are subject to the $\widehat{U(1)}$ Kac-Moody algebra relation,
\begin{equation}\label{lyp23}
\left[j_{m}, j_{n}\right]=km\delta_{m+n,0},
\end{equation}
and $k$ is level of the representation. So \eqref{lyp21} is a realization of current algebra $\widehat{U(1)}_{k}$. The bosonic current is
\begin{equation}\label{lyp22}
J(z)=\sum_{n\in \mathbb{Z}}j_{n}z^{-n-1}.
\end{equation}
From it, the energy momentum tensor is given as
\begin{equation}\label{lyp24}
L(z)=\dfrac{1}{2k}:J(z)J(z):=\sum_{n\in\mathbb{Z}}L_{n}z^{-n-2}.
\end{equation}
Here, the modes $L_{n}$ are
\begin{equation}\label{lyp25}
L_{n}=\dfrac{1}{2k}\left\{\dfrac{k^{2}}{N^{2}}\sum_{l=0}^{n}\dfrac{\partial^{2}}{\partial t_{l}\partial t_{n-l}}+2k\sum_{l=1}^{\infty}lt_{l}\dfrac{\partial}{\partial t_{l+n}}\right\},\ \ n\geq-1,
\end{equation}
and
\begin{equation}\label{lyp26}
L_{n}=\dfrac{1}{2k}\left\{2k\sum_{l=0}^{\infty}(-n+l)t_{-n+l}\dfrac{\partial}{\partial t_{l}}+N^{2}\sum_{l=1}^{-n-1}l(-n-l)t_{l}t_{-n-l}\right\},\ \ n\leq-2.
\end{equation}

From the above expression, it is easy to see that when the level $k$ is restricted to 2, the annihilation subalgebra (i.e. $\{L_{n}, n\geq-1\}$) is exactly the Virasoro constraint of one-hermitian matrix model. In the case $k=2$, denoting the current \eqref{lyp24} by $\hat{L}(z)$, then
\begin{equation}\label{lyp37}
\begin{array}{llll}
\hat{L}(z)&=&\dfrac{1}{4}:\left(\sum_{n=0}^{\infty}\dfrac{2}{N}\dfrac{\partial}{\partial t_{n}}z^{-n-1}+\sum_{n=1}^{\infty}Nnt_{n}z^{n-1}\right)^{2}:\\
&=&\sum\limits_{n\in\mathbb{Z}}\hat{L}_{n}z^{-n-2}.
\end{array}
\end{equation}
So the explicit expression of $\hat{L}_{n}$ are
\begin{equation}\label{lyp38}
\hat{L}_{n}=\dfrac{1}{N^{2}}\sum_{k=0}^{n}\dfrac{\partial^{2}}{\partial t_{k}\partial t_{n-k}}+\sum_{k=1}^{\infty}kt_{k}\dfrac{\partial}{\partial t_{n+k}},\ \ n\geq-1,
\end{equation}
and
\begin{equation}\label{lyp39}
\hat{L}_{n}=\sum_{l=0}^{\infty}(-n+l)t_{-n+l}\dfrac{\partial}{\partial t_{l}}+\dfrac{N^{2}}{4}\sum_{l=1}^{-n-1}l(-n-l)t_{l}t_{-n-l},\ \ n\leq-2.
\end{equation}
The set of operators $\{\hat{L}_{n}; n\in\mathbb{Z}\}$ is a realization of Virasoro algebra with central charge $c=1$,
\begin{equation}\label{lyp27}
\left[\hat{L}_{n}, \hat{L}_{m}\right]=(n-m)\hat{L}_{m+n}+\dfrac{1}{12}(n^{3}-n)\delta_{n+m,0}.
\end{equation}

\section{Generalized Virasoro Constraints}
In section 3, we have given the Virasoro constraint of one-hermitian matrix model, i.e. the operator representation of equation \eqref{lyp10}. We now derive a more general constraint of this matrix model.

Using the expression of $V(x)$, expand $\overline{W}_{k}(x_{1},\cdots,x_{k})$ in equation \eqref{lyp11} as \eqref{lyp16} with respect the parameters $x_{1},\cdots,x_{k}$
\begin{equation}\label{lyp28}
\begin{array}{lllll}
\sum\limits_{n=-1}^{\infty}\sum\limits_{n_{1}=1}^{\infty}\cdots
\sum\limits_{n_{k=1}}^{\infty}x^{-n-2}x_{1}^{-n_{1}-1}\cdots x_{k}^{-n_{k}-1}\\
\cdot\left\{N^{k}\sum\limits_{l=0}^{n}\left\langle trM^{l}trM^{n-l}trM^{n_{1}}trM^{n_2}\cdots trM^{n_{k}}\right\rangle\right.\\
\left.-N^{k+1}\sum\limits_{l=1}^{\infty}lt_{l}\left\langle trM^{n+l}trM^{n_{1}}trM^{n_{2}}\cdots trM^{n_{k}}\right\rangle\right.\\
\left.+N^{k}\sum\limits_{l=1}^{k}n_{l}\langle trM^{n_{l}+n}trM^{n_{1}}\cdots \widehat{trM^{n_{l}}}\cdots trM^{n_{k}}\rangle\right\}=0.
\end{array}
\end{equation}
Similarly as equation \eqref{lyp18}, equation \eqref{lyp28} can also be reexpressed as operators acting on the partition function $Z$:
\begin{equation}\label{lyp29}
\begin{array}{lllll}
\sum\limits_{n=-1}^{\infty}\sum\limits_{n_{1}=1}^{\infty}\cdots
\sum\limits_{n_{k=1}}^{\infty}x^{-n-2}x_{1}^{-n_{1}-1}\cdots x_{k}^{-n_{k}-1}\cdot\dfrac{(-1)^{k}}{Z}\cdot\\
\cdot\left\{\sum\limits_{l=0}^{n}\dfrac{1}{N^{2}}\dfrac{\partial}{\partial t_{l}}\dfrac{\partial}{\partial t_{n-l}}\dfrac{\partial}{\partial t_{n_{1}}}\cdots\dfrac{\partial}{\partial t_{n_{k}}}+\sum\limits_{l=1}^{\infty}lt_{l}\dfrac{\partial}{\partial t_{n+l}}\dfrac{\partial}{\partial t_{n_{1}}}\cdots\dfrac{\partial}{\partial t_{n_{k}}}\right.\\
\left.+\sum\limits_{l=1}^{k}n_{l}\dfrac{\partial}{\partial t_{n_{l}+n}}\dfrac{\partial}{\partial t_{n_{1}}}\cdots\hat{\dfrac{\partial}{\partial t_{n_{l}}}}\cdots\dfrac{\partial}{\partial t_{n_{k}}}\right\}\cdot Z=0,
\end{array}
\end{equation}
in the above equations, " $\hat{}$ " means that the corresponding term will not appear in the series.

We use the standard notation as in the book \cite{Macdonald}. One says $\lambda$ is a partition, it means\\ $\lambda=(\lambda_{1}, \lambda_{2},\cdots,\lambda_{l}),\  \lambda_{1}\geq\lambda_{2}\geq\cdots\geq\lambda_{l}>0$. The numbers of the non-zero $\lambda_{i}$ is the length of the partition $\lambda$, denoted as $l(\lambda)$ and the weight of partition $\lambda$ is
\begin{equation}\label{lyp30}
|\lambda|=\sum_{i=1}^{l(\lambda)}\lambda_{i}.
\end{equation}
We denote the set of all partitions by $\mathscr{P}$. If $\lambda=(\lambda_{1},\cdots,\lambda_{l(\lambda)})$ and $\mu=(\mu_{1},\cdots,\mu_{l(\mu)})$ are two partitions, we define the following operations between them,
\begin{equation}\label{lyp34}
\begin{array}{lll}
\lambda\cup\mu=(\nu_{1},\nu_{2},\cdots,\nu_{l(\lambda)+l(\mu)}), \ \nu_{1}\geq\nu_{2}\geq\cdots\geq\nu_{l(\lambda)+l(\mu)},\\
\nu_{i}\in\lambda\  \text{or}\   \nu_{i}\in\mu,\ \ \text{for}\   1\leq i\leq l(\lambda)+l(\mu),
\end{array}
\end{equation}
and
\begin{equation}\label{lyp35}
\lambda\backslash \lambda_{i}=(\lambda_{1},\cdots,\lambda_{i-1},\lambda_{i+1},\cdots,\lambda_{l(\lambda)}).
\end{equation}
Now, we can introduce operators
\begin{equation}\label{lyp31}
\begin{array}{llll}
\mathscr{L}_{n;\lambda}&=&\sum\limits_{k=0}^{n}\dfrac{1}{N^{2}}\dfrac{\partial}{\partial t_{k}}\dfrac{\partial}{\partial t_{n-k}}\prod_{i=1}^{l(\lambda)}\dfrac{\partial}{\partial t_{\lambda_{i}}}+\sum_{k=1}^{\infty}kt_{k}\dfrac{\partial}{\partial t_{n+k}}\prod_{i=1}^{l(\lambda)}\dfrac{\partial}{\partial t_{\lambda_{i}}}\\
&&+\sum\limits_{i=1}^{l(\lambda)}\lambda_{i}\dfrac{\partial}{\partial t_{\lambda_{i}+n}}\prod_{j=1,j\neq i}^{l(\lambda)}\dfrac{\partial}{\partial t_{\lambda_{j}}}.
\end{array}
\end{equation}
Equation \eqref{lyp29} is equivalent to an equation be expressed through the operators $\mathscr{L}_{n;\lambda}$
\begin{equation}\label{lyp58}
\sum\limits_{n=-1}^{\infty}\sum\limits_{\lambda\in\mathscr{P}}x^{-n-2}x_{1}\cdots x_{k}m_{\lambda}(x_{1}^{-1},\cdots,x_{k}^{-1})\mathscr{L}_{n;\lambda}\cdot Z=0
\end{equation}
Here $m_{\lambda}(x_{1}^{-1},\cdots,x_{k}^{-1})$ is the monomial symmetric function generated by $x_{1}^{-\lambda_{1}}\cdots x_{k}^{-\lambda_{k}}$.
\begin{equation}\label{lyp62}
m_{\lambda}(x_{1}^{-1},\cdots,x_{k}^{-1})=\sum\limits_{\alpha}x_{1}^{-\alpha_{1}}\cdots x_{k}^{-\alpha_{k}},
\end{equation}
with summing over all distinct permutations $(\alpha_{1},\cdots,\alpha_{k})$ of $\lambda=(\lambda_{1},\cdots,\lambda_{k})$.
In equations \eqref{lyp29} and \eqref{lyp58}, $k$ could be any positive integer. When $k\rightarrow\infty$, the monomial symmetric functions  $m_{\lambda}(x_{1}^{-1},\cdots,x_{k}^{-1})$, $\lambda\in\mathscr{P}$ are independent. So all the coefficients of  $x^{-n-2}x_{1}\cdots x_{k}m_{\lambda}(x_{1}^{-1},\cdots,x_{k}^{-1})$ in equation \eqref{lyp58} are zero. That is
\begin{equation}\label{lyp32}
\mathscr{L}_{n;\lambda}\cdot Z=0,  n\geq-1.
\end{equation}
The set of operators $\{\mathscr{L}_{n;\lambda}, n\geq-1\}$ forms a constraint of the one-hermitian matrix model \eqref{lyp1}. The Virasoro constraint $\{\mathscr{L}_{n},n\geq-1\}$ introduced in section $3$ is a spacial case of $\{\mathscr{L}_{n;\lambda}, n\geq-1\}$ with $\lambda=0$. We could name $\{\mathscr{L}_{n;\lambda}, n\geq-1\}$ as a generalized Virasoro constraint of 1-hermitian matrix model, for they satisfy the following relation:
\begin{equation}\label{lyp33}
\begin{array}{lll}
&&\left[\mathscr{L}_{n;\lambda}, \mathscr{L}_{m;\mu}\right]\\
=&&(n-m)\mathscr{L}_{n+m;\lambda\cup\mu}\\
&&+\sum\limits_{i=1}^{l(\lambda)}\lambda_{i}\mathscr{L}_{n;\lambda\cup\mu\cup(m+\lambda_{i})
\backslash\lambda_{i}}\\
&&-\sum\limits_{j=1}^{l(\mu)}\mu_{j}\mathscr{L}_{m;\lambda\cup\mu\cup(n+\mu_{j})\backslash\mu_{j}}.
\end{array}
\end{equation}
In equation \eqref{lyp33}, if both the partitions $\lambda$ and $\mu$ are $\emptyset$, it is the usual Virasoro constraint \eqref{lyp20}. It is easy to verify that the Lie bracket defined in equation \eqref{lyp33} satisfies
\begin{equation}\label{lyp59}
\left[\mathscr{L}_{n;\lambda},\mathscr{L}_{m;\mu}\right]=-\left[\mathscr{L}_{m;\mu},\mathscr{L}_{n;\lambda}\right]
\end{equation}
and the Jacobi identity
\begin{equation}\label{lyp60}
\left[\mathscr{L}_{m;\lambda},\left[\mathscr{L}_{n;\mu},\mathscr{L}_{k;\nu}\right]\right]+
\left[\mathscr{L}_{n;\mu},\left[\mathscr{L}_{k;\nu},\mathscr{L}_{m;\lambda}\right]\right]+
\left[\mathscr{L}_{k;\nu},\left[\mathscr{L}_{m;\lambda},\mathscr{L}_{n;\mu}\right]\right]=0,
\end{equation}
where $m,n$ and $k$ are integers bigger than $-2$ and $\lambda,\mu$ and $\mu$ are partitions. These imply that $\{\mathscr{L}_{n;\lambda},n\leq-1\}$ forms a Lie algebra. This algebra include more information than the annihilation subalgebra of the Virasoro algebra. Even for the case $m=n=0$ in \eqref{lyp33},
\begin{equation}\label{lyp61}
\begin{array}{lll}
&&\left[\mathscr{L}_{0;\lambda}, \mathscr{L}_{0;\mu}\right]\\
=&&\sum\limits_{i=1}^{l(\lambda)}\lambda_{i}\mathscr{L}_{0;\lambda\cup\mu\cup(m+\lambda_{i})
\backslash\lambda_{i}}-\sum\limits_{j=1}^{l(\mu)}\mu_{j}\mathscr{L}_{0;\lambda\cup\mu\cup(n+\mu_{j})\backslash\mu_{j}}.
\end{array}
\end{equation}
$\{\mathscr{L}_{0;\lambda}\}$ is a nontrivial subalgebra of $\{\mathscr{L}_{n;\lambda}\}$.
Equation \eqref{lyp31} gives the definition of operators $\mathscr{L}_{n;\lambda}$ with $n\geq-1$.

Now we will extend the set of operators$\{\mathscr{L}_{n;\lambda},n\leq-1\}$ in a way to $\{\hat{L}_{\Lambda;\mu,\lambda}\}$, such that $\{\hat{L}_{\Lambda;\mu,\lambda}\}$ forms a Lie algebra, and $\{\mathscr{L}_{n;\lambda}\}$ for $n\geq-1$ is a subalgebra of the extended algebra.

From section 3, the set of operators $\{\hat{L}_{n};n\in\mathbb{Z}\}$ forms a Virasoro algebra and $\hat{L}_{n}=\mathscr{L}_{n}$ for $n\geq-1$. If
\begin{equation}\label{lyp36}
\hat{j}_{n}=\left\{\begin{array}{lll}j_{n}=\dfrac{2}{N}\dfrac{\partial}{\partial t_{n}}, n>0;\\
j_{n}=-Nnt_{-n},  n\leq0;
\end{array}\right.,
\end{equation}
and then the generating function of these operators is
\begin{equation}\label{lyp42}
\hat{J}(z)=\sum_{n\in\mathbb{Z}}\hat{j}_{n}z^{-n-1}.
\end{equation}
One easily gets
\begin{equation}\label{lyp45}
\left[\hat{L}_{n}, \hat{j}_{k}\right]=-k\hat{j}_{n+k}.
\end{equation}
Let $\lambda=(\lambda_{1},\cdots,\lambda_{l(\lambda)})$ be a partition of positive integer, we set the following operators indexed by the partitions:
\begin{equation}\label{lyp40}
\hat{j}_{\lambda}=\hat{j}_{\lambda_{1}}\hat{j}_{\lambda_{2}}\cdots\hat{j}_{l(\lambda)},
\end{equation}
and
\begin{equation}\label{lyp41}
\hat{j}_{\lambda}^{\dag}=\hat{j}_{-\lambda_{1}}\hat{j}_{-\lambda_{2}}\cdots\hat{j}_{-l(\lambda)}.
\end{equation}
Now we make a new notation: generalized partitions. Let $\Lambda$ be a generalized partition, it means:
\begin{equation}\label{lyp43}
\begin{array}{lll}
\Lambda=(\Lambda_{1},\Lambda_{2},\cdots,\Lambda_{l(\Lambda)}),\\
-\infty<\Lambda_{1}\leq\Lambda_{2}\leq\cdots\leq\Lambda_{l(\Lambda)}<\infty;\ \text{and} \
  \Lambda_{i}\in\mathbb{Z}\ \text{for}\  1\leq i\leq l(\Lambda).
\end{array}
\end{equation}
Similarly, we denote $l(\Lambda)$ as the length of generalized partition $\Lambda$, and the weight of $\Lambda$ is the algebraic summation of $\Lambda_{i}$
\begin{equation}\label{lyp50}
|\Lambda|=\sum\limits_{i=1}^{l(\Lambda)}\Lambda_{i}.
\end{equation}
For a generalized partition $\Lambda$, we set
\begin{equation}\label{lyp44}
\hat{L}_{\Lambda}=\hat{L}_{\Lambda_{1}}\hat{L}_{\Lambda_{2}}\cdots\hat{L}_{\Lambda_{l(\Lambda)}}.
\end{equation}
As well as partitions $\lambda$ and $\mu$,
\begin{equation}\label{lyp46}
\hat{L}_{\Lambda;\mu,\lambda}=\hat{j}_{\mu}^\dag\hat{j}_{\lambda}\hat{L}_{\Lambda},
\end{equation}
and especially
\begin{equation}\label{lyp47}
\hat{L}_{\emptyset;\emptyset,\emptyset}=1.
\end{equation}
It is easily to see that the set of operators $\{\hat{L}_{\Lambda;\mu,\lambda}\}$ forms a Lie algebra, i.e.
\begin{equation}\label{lyp48}
\begin{array}{lllll}
\left[\hat{L}_{\Lambda;\mu,\lambda}, \hat{L}_{\Lambda^{\prime};\mu^{\prime},\lambda^{\prime}}\right]=\sum\limits_{\Lambda^{\prime\prime};
\mu^{\prime\prime},\lambda^{\prime\prime}}
c_{\Lambda^{\prime\prime};\mu^{\prime\prime},\lambda^{\prime\prime}}\hat{L}_{\Lambda^{\prime\prime};
\mu^{\prime\prime},\lambda^{\prime\prime}};\\
\left[\hat{L}_{\Lambda;\mu,\lambda}, \hat{L}_{\Lambda^{\prime};\mu^{\prime},\lambda^{\prime}}\right]=-\left[\hat{L}_{\Lambda^{\prime};
\mu^{\prime},\lambda^{\prime}}, \hat{L}_{\Lambda;\mu,\lambda}\right];\\
\left[\left[\hat{L}_{\Lambda;\mu,\lambda}, \hat{L}_{\Lambda^{\prime};\mu^{\prime},\lambda^{\prime}}\right], \hat{L}_{\Lambda^{\prime\prime};\mu^{\prime\prime},\lambda^{\prime\prime}}\right]+
\left[\left[\hat{L}_{\Lambda^{\prime};\mu^{\prime},\lambda^{\prime}}, \hat{L}_{\Lambda^{\prime\prime};\mu^{\prime\prime},\lambda^{\prime\prime}}\right], \hat{L}_{\Lambda;\mu,\lambda}\right]\\
+\left[\left[\hat{L}_{\Lambda^{\prime\prime};\mu^{\prime\prime},\lambda^{\prime\prime}}, \hat{L}_{\Lambda;\mu,\lambda}\right], \hat{L}_{\Lambda^{\prime};\mu^{\prime},\lambda^{\prime}}\right]=0,
\end{array}
\end{equation}
where $c_{\Lambda^{\prime\prime};\mu^{\prime\prime};\lambda^{\prime\prime}}$ are structure constants, $\Lambda$, $\Lambda^{\prime}$ and $\Lambda^{\prime\prime}$ are arbitrary generalized partitions, while $\lambda$, $\lambda^{\prime}$, $\lambda^{\prime\prime}$, $\mu$, $\mu^{\prime}$ and $\mu^{\prime\prime}$ are arbitrary partitions of positive integers, respectively. For explicit partitions, we could give the exact value of the structure constant $c$. But for general partitions, it is very complicated and could be not indexed by partitions simply. On the other hand, the structure constants $c$ should rely on a definite representation of the algebra. Besides the bracket $[\hat{L}_{\Lambda;\lambda,\mu},\hat{L}_{\Lambda^{\prime};\lambda^{\prime},\mu^{\prime}}]$, we can make some other brackets, such as
\begin{equation}\label{lyp63}
\begin{array}{llllllll}
\left[\hat{L}_{\Lambda;\lambda,\mu},\hat{L}_{\Lambda^{\prime};\lambda^{\prime},\mu^{\prime}}\right]_{1}=
\hat{L}_{\Lambda;\lambda,\mu}\hat{L}_{\Lambda^{\prime};\lambda^{\prime},\mu^{\prime}}-
\hat{L}_{\Lambda^{\prime};\lambda,\mu}\hat{L}_{\Lambda;\lambda^{\prime},\mu^{\prime}},\\
\left[\hat{L}_{\Lambda;\lambda,\mu},\hat{L}_{\Lambda^{\prime};\lambda^{\prime},\mu^{\prime}}\right]_{2}=
\hat{L}_{\Lambda;\lambda,\mu}\hat{L}_{\Lambda^{\prime};\lambda^{\prime},\mu^{\prime}}-
\hat{L}_{\Lambda;\lambda^{\prime},\mu}\hat{L}_{\Lambda^{\prime};\lambda,\mu^{\prime}},\\
\left[\hat{L}_{\Lambda;\lambda,\mu},\hat{L}_{\Lambda^{\prime};\lambda^{\prime},\mu^{\prime}}\right]_{3}=
\hat{L}_{\Lambda;\lambda,\mu}\hat{L}_{\Lambda^{\prime};\lambda^{\prime},\mu^{\prime}}-
\hat{L}_{\Lambda;\lambda,\mu^{\prime}}\hat{L}_{\Lambda^{\prime};\lambda^{\prime},\mu},\\
\left[\hat{L}_{\Lambda;\lambda,\mu},\hat{L}_{\Lambda^{\prime};\lambda^{\prime},\mu^{\prime}}\right]_{4}=
\hat{L}_{\Lambda;\lambda,\mu}\hat{L}_{\Lambda^{\prime};\lambda^{\prime},\mu^{\prime}}-
\hat{L}_{\Lambda;\lambda^{\prime},\mu^{\prime}}\hat{L}_{\Lambda^{\prime};\lambda,\mu},\\
\left[\hat{L}_{\Lambda;\lambda,\mu},\hat{L}_{\Lambda^{\prime};\lambda^{\prime},\mu^{\prime}}\right]_{5}=
\hat{L}_{\Lambda;\lambda,\mu}\hat{L}_{\Lambda^{\prime};\lambda^{\prime},\mu^{\prime}}-
\hat{L}_{\Lambda^{\prime};\lambda^{\prime},\mu}\hat{L}_{\Lambda;\lambda,\mu^{\prime}},\\
\left[\hat{L}_{\Lambda;\lambda,\mu},\hat{L}_{\Lambda^{\prime};\lambda^{\prime},\mu^{\prime}}\right]_{6}=
\hat{L}_{\Lambda;\lambda,\mu}\hat{L}_{\Lambda^{\prime};\lambda^{\prime},\mu^{\prime}}-
\hat{L}_{\Lambda^{\prime};\lambda,\mu^{\prime}}\hat{L}_{\Lambda;\lambda^{\prime},\mu}.
\end{array}
\end{equation}
It is easy to verify that $[\hat{L}_{\Lambda;\lambda,\mu},\hat{L}_{\Lambda^{\prime};\lambda^{\prime},\mu^{\prime}}]_{i}, i=1,2\cdots,6$ belong to $\{\hat{L}_{\Lambda;\lambda,\mu}\}$ too. The explicit expressions of them are very complicated,  and these brackets may not have straightforward relationship with matrix model, we will not discuss them in detail here.

For $\Lambda=(n),n\geq-1$ and $\mu=0$, then
\begin{equation}
\hat{L}_{(n);0,\lambda}=\left(\dfrac{2}{N}\right)^{l(\lambda)}\mathscr{L}_{n,\lambda}.
\end{equation}
So the generalized Virasoro constraints $\{\mathscr{L}_{n}; n\geq-1\}$ is a subalgebra of operator algebra $\{\hat{L}_{\Lambda;\mu,\lambda}\}$, and it also gives a bosonic realization to the generalized Virasoro constraints of one-hermitian matrix model.

From the algebraic relation \eqref{lyp33}, we could also obtain an algebra $\mathcal {L}$.
The elements of algebra $\mathcal {L}$ are $\mathcal {L}_{n,\Lambda}$ for $n\in\mathbb{Z}$, and $\Lambda$ is any generalized partition. They satisfy the following relation:
\begin{equation}\label{lyp49}
\begin{array}{llllll}
&&\left[\mathcal {L}_{n,\Lambda}, \mathcal {L}_{m,\Lambda^{\prime}}\right]\\
=&&\left(n-m\right)\mathcal {L}_{n+m,\Lambda\cup\Lambda_{\prime}}\\
&&+\sum\limits_{i=1}^{l(\Lambda)}\Lambda_{i}\mathcal {L}_{n,\Lambda\cup\Lambda^{\prime}\cup(\Lambda_{i}+m)\backslash\Lambda_{i}}\\
&&+\sum\limits_{j=1}^{l(\Lambda^{\prime})}\Lambda^{\prime}_{j}\mathcal {L}_{m,\Lambda\cup\Lambda\cup(\Lambda^{\prime}_{j}+n)\backslash\Lambda^{\prime}_{j}}\\
&&+\dfrac{c}{12}\left(n+|\Lambda|+1\right)\left(n+|\Lambda|\right)\left(n+|\Lambda|-1\right)
\delta_{n+m+|\Lambda|+|\Lambda^{\prime}|,0}.
\end{array}
\end{equation}
where $c$ is a complex number.
It is easy to verify the Lie bracket:
\begin{equation}\label{lyp51}
\begin{array}{lllll}
\left[\mathcal {L}_{n,\Lambda}, \mathcal {L}_{m,\Lambda^{\prime}}\right]=-\left[\mathcal {L}_{m,\Lambda^{\prime}}, \mathcal {L}_{n,\Lambda}\right],
\end{array}
\end{equation}
and the Jocobi identity:
\begin{equation}\label{lyp53}
\begin{array}{llllll}
\left[\left[\mathcal {L}_{n,\Lambda}, \mathcal {L}_{m,\Lambda^{\prime}}\right], \mathcal {L}_{k,\Lambda^{\prime\prime}}\right]+\left[\left[\mathcal {L}_{m,\Lambda^{\prime}}, \mathcal {L}_{k,\Lambda^{\prime\prime}}\right], \mathcal {L}_{n,\Lambda}\right]\\
+\left[\left[\mathcal {L}_{k,\Lambda^{\prime\prime}}, \mathcal {L}_{n,\Lambda}\right], \mathcal {L}_{m,\Lambda^{\prime}}\right]=0,
\end{array}
\end{equation}
respectively. Here $n$, $m$ and $k$ are arbitrary integers and $\Lambda$, $\Lambda^{\prime}$ and $\Lambda^{\prime\prime}$ are arbitrary generalized partitions, respectively. From equation \eqref{lyp49}, and the Lie bracket \eqref{lyp51}, as well as the Jacobi identity \eqref{lyp53}, we see that $\mathcal {L}$ is a  Lie algebra, and it is a generalization of the usual Virasoro algebra. If we fix all the generalized partition as $\emptyset$ in \eqref{lyp49}, it is exactly the usual Virasoro algebra. The algebra $\mathcal {L}$ contains more information than the Virasoro algebra. Even $n=m=0$ in equation \eqref{lyp49}, $\{\mathcal {L}_{0;\Lambda}\}$
forms a nontrivial subalgebra of $\mathcal {L}$, while when $n=m=0$ in the Virasoro algebra, it is trivial. In fact, we could have another two kinds brackets for $\{\mathcal {L}_{n;\Lambda}\}$,  these brackets may not have straightforward relationship with matrix model, so we do not list them here.

The set $\{\mathcal {L}_{n,\Lambda}; n\geq-1, \Lambda_{i}>0\}$ forms a subalgebra of $\mathcal {L}$ under the relation \eqref{lyp49}, and it is exactly the generalized Virasoro constraint. From equation \eqref{lyp32}, one can easily find that the Hermitian one-matrix model forms a natural representation of this subalgebra. Like the Virasoro algebra, there must be some interesting representations for the algebra $\mathcal {L}$, such as the highest weight representation etc. To investigate the subject is one of our aim in the future.

It is nature to expect that, the algebra $\mathcal {L}$ should correspond to certain field theory. As it is known that, the two-dimensional Wess-Zumino--Novikov-Witten (WZNW) model is the highest weight representation of affine Lie (Kac-Moody) algebras, and the Virasoro algebra corresponds to the model is obtained through the Sugawara construction. As the algebra $\mathcal {L}$ contains more information than the Virasoro algebra, so the field theory corresponding to the new algebra should be more complicated than the WZNW model, and the field theory should be an extension of the well known WZNW model . To find such model is our another goal in the future. We hope the algebra $\mathcal {L}$ may has relationship with the $4D$ $N=2$ supersymmetric Yang-Mills theory. If this point view is right, one could get certain correspondence between different field theories from this algebra.

\section{Conclusion}
In this article,  from the standard CFT method we give a realization of Virasoro algebra whose annihilation subalgebra is exactly the Virasoro constraint of the Hermitian one-matrix model. Similar as the equivalence between the one-loop equation and the Virasoro constraint, we get a generalized Virasoro constraint from multi-loop equations of the Hermitian one-matrix model. Extending the generalized Virasoro constraint, we obtain more interesting algebras. $\{\hat{L}_{\Lambda;\mu,\lambda}\}$ is an extension of the generalized Virasoro constraint. In fact, the algebra $\{\hat{L}_{\Lambda;\mu,\lambda}\}$ is an universal enveloping algebra of the algebra $\{\hat{j}_{m},\hat{L}_{n};m,n\in\mathbb{Z}\}$. And $\{\hat{L}_{\Lambda;\mu,\lambda}\}$ provides a bosonic realization to the generalized Virasoro algebra. A more interesting extension is the algebra $\mathcal {L}$, as a Lie bracket of $\mathcal {L}$ is very natural, and its the elements are indexed by partitions. The algebra $\mathcal {L}$ should be universal in the subjects related to matrix model. We want to find a nontrivial realization of $\mathcal {L}$, but we could not get it up to now.

For any algebraic curve, B.Eynard and N.Orantin have constructed a sequence of symplectic invariants on it. Particularly, the partition function of the Hermitian one-matrix model is exactly the symplectic invariant on its spectral curve. They have also introduced Virasoro constraints of these symplectic invariants locally and globally on the spectral curve. The origin of these Virasoro constraints is the one for the  Hermitian matrix model. On any curve, there should be certain constraints corresponding to the generalized Virasoro constraint and the algebra $\mathcal {L}$ we have presented here. To find the correspondence is an interesting issue for us.

\section*{Acknowledgments}

The financial supports from the National Natural Science Foundation
of China (Grant No. 11375258) are gratefully
acknowledged from one of the author (Ding). The authors would like
to thank the Kavli Institute for Theoretical Physics China at the
Chinese Academy of Sciences, where part of the work was done in the
period of the program entitled "Mathematical Methods from Physics" hold at July 22--Sept. 5, 2013.

\end{document}